\documentclass[12pt, a4paper,twoside,useAMS]{article}
\usepackage{epsfig}

\usepackage{baltlat2}

\usepackage{wrapfig}

\def\lag{{\mbox{${\cal L}$}} {}}

\newcommand{\rmd}{{\rm d}}
\begin{document}
\ \

\vspace{0.5mm}

\setcounter{page}{0}

\vspace{8mm}

\titlehead{Baltic Astronomy, vol.\ts 14, 000-000, 2005.}

\titleb{LIMITATIONS OF THE HAMILTONIAN TREATMENT\\ FOR 
COLLISIONLESS ASTROPHYSICAL ACCRETION FLOWS}

\begin{authorl}
\authorb{Vladimir~I.~Pariev}{1,2} and
\authorb{Eric~G.~Blackman}{3}
\end{authorl}

\begin{addressl}
\addressb{1}{Physics Department, University of Wisconsin-Madison,\\
1150 University Avenue, Madison, WI 53706, U.S.A.}

\addressb{2}{P.N. Lebedev Physical
Institute, Leninsky Prospect 53,\\ Moscow 119991, Russia}

\addressb{3}{Department of Physics and Astronomy, University of Rochester,\\
Rochester, NY 14627, U.S.A.}
\end{addressl}

\submitb{Received ..., 2005}

\begin{abstract} 
In the Hamiltonian treatment of  purely mechanical systems,
the canonical and actual momentum of a particle are the same.
In contrast, for a plasma of charged particles and electromagnetic
fields, those two momenta are different.
We show how this distinction is fundamental in identifying the
limitations of a recent attempt by Binney (2003) to rule out two-temperature
collisionless astrophysical accretion flows from Hamiltonian theory.
This illustrates the Hamiltonian method for astrophysical plasmas, 
its relation to the equations of motion, and its role in practical calculations.
We also discuss how the complete Hamiltonian treatment of a plasma 
should couple the particle
motion to a fully dynamical treatment of the  electromagnetic fields.
Our results stand independent from the discussion of Quataert (2003)
who argued that time scale calculated in Binney (2003)
is not the equipartition time as claimed. 
\end{abstract}


\resthead{Limitations of the Hamiltonian treatment}{Vladimir I. Pariev,
Eric G. Blackman}

\sectionb{1}{INTRODUCTION}\label{sec1} 

An accretion disc of hot plasma orbiting a massive black hole 
and slowly spiralling in from the action of viscosity is widely considered
to be the main energy source for very luminous extragalactic objects such as 
quasars and other anomalously bright active galactic nuclei (AGN)
(Shapiro \& Teukolsky 1983; Krolik 1999). Recently, the determination of 
the density and temperature of the gas in the accretion disc
surrounding the central black hole of  
our own Milky Way Galaxy as well as discs around other  nearby galactic 
centres has become possible from X-ray  
data taken by orbiting X-ray telescopes such as XMM Newton
and the Chandra X-ray Observatory. 
From these measurements, the  accretion rate, ${\dot M}$, can be 
determined.  The standard theory of geometrically thin accretion discs 
around black holes predicts that about $0.1$  of ${\dot M} c^2$ 
is converted into escaping radiation. This estimate is consistent with the 
radiative efficiency of the integrated luminosity 
of quasars with the observed space density of supermassive black holes.
However, a puzzling discrepancy has emerged:
the measured luminosities  of the central sources in some nearby galaxies 
are smaller than this standard estimate by $3$ to $5$ orders of magnitude.

This discrepancy 
led to the development of a newer, but also popular, 
type of geometrically thick 
accretion disc models called ``advection dominated accretion
flows'' (ADAFs) ((e.g., Ichimaru 1977; Rees et al. 1982
Narayan \& Yi 1995; Narayan, Mahadevan \& Quataert 1998)
in which the gravitational binding energy of the accreting
material is retained as 
internal energy within the hot plasma and ultimately crosses the event horizon
of a black hole as the plasma falls in without radiating significantly. 
In the vicinity of the black hole horizon, the gravitational
binding energy of the plasma is a fraction of ${\dot M} c^2$ and, therefore,
the associated internal energy per particle in an ADAF 
is of order $m_{\rm p} c^2 \sim 1\,\mbox{GeV}$. 
Because electrons are more mobile, they are the primary radiating
particles.  Thus, in order for such a weakly radiating accretion flow to exist, 
three main assumptions of the ADAF model must be satisfied: 
(1)  the internal energy dissipated in the accretion process via
viscosity must go almost entirely into ions.
(2) the heat transfer from ions to electrons must be slow enough, 
so that only a tiny fraction of the dissipated 
thermal energy received by the ions is transferred to electrons 
during the time it takes the gas to lose its angular momentum and fall onto
the black hole. 
(3) the effective viscosity must be very high in order
that the gas looses its angular
momentum quickly and can indeed accrete faster than the ion-electron
thermal coupling times.

If the assumptions were true,
an accretion flow with a given ${\dot M}$ could have a low
enough number density and high enough temperature 
that Coulomb collisions 
are inefficient in establishing equipartition of energy between ions and 
electrons during the accretion time. 
In the absence of any other plasma process that could  speed up 
the ion-electron energy transfer, the electrons could remain at 
temperatures which are 
orders of magnitude lower than the ions in the black hole engine environments. 
Since electrons produce 
practically all of the radiation in a given disc, 
the luminosity rises steeply
with increasing  electron temperature. Thus 
keeping electrons 
at a much lower temperature achieves the main goal of ADAFs:
substantially lowering the luminosity 
for a given accretion rate 
(e.g., Ichimaru 1977;
Narayan \& Yi 1995; 
Quataert \& Gruzinov 2000; Narayan, Igumenshchev
\& Abramowicz 2000; Narayan 2002) compared to standard thin discs.
ADAFs are thick discs because the heat dissipated by the accretion
is stored as thermal energy of the ions, which puffs up the disc.

Despite the important implications of the above assumptions
if they were true, these assumptions have not been proven or disproven.
Doing so requires understanding the  subtle plasma physics of
the interactions between ions and electrons  with magnetic and electric fields. The assumptions have therefore been the subject of much deserved attention
(Begelman \& Chiueh 1988; 
Bitsnovatyi-Kogan \& Lovelace 1997; Quataert 1998; Gruzinov 1998; 
Blackman 1999; Quataert \& Gruzinov 2000).
One central issue is whether or not 
 collective long range interactions could be  important for momentum
transfer, energy dissipation and thermal equilibration processes occurring 
not just between pairs of particles (like Coulomb collisions) but in the whole 
volume of plasma, shortening the electron ion equilibration time and
ruling out the ADAFs.

Since the basic microphysical 
processes of a plasma must involve known electromagnetic  interactions 
between particles and fields, 
it is tempting to address the above ADAF assumptions 
starting from a very basic treatment of 
electromagnetic theory. This was 
recently pursued  by Binney (2003),
who presented a general argument that assumption (2) above 
is invalid. He used  a  Hamiltonian 
formalism to calculate the change of energy and  
angular momentum of a particle moving in a general 
time-dependent electromagnetic
field and averaged the result obtained over 
many particles to estimate the corresponding rates for the plasma. 
He concluded that the ion-electron 
equipartition time $t_{\rm equi}$ is smaller than the characteristic 
angular momentum loss time, $t_{\rm res}$ (or residence
time according to Binney's (2003) terminology). 
If this result were true, and the approach correct,
it would rule out two-temperature accretion flows and close the whole
ADAF chapter in astrophysics.

Scientists working in plasma astrophysics usually start with writing
down Lorentz forces acting on particles and Maxwell equations to 
analyse interactions in astrophysical plasmas. The idea of Binney 
(2003) to invoke general Hamiltonian analysis is novel and 
original. 
However, in this paper we follow along with the calculation of Binney (2003)
and discover two fatal problems therein:
(1) The calculation of Binney assumed that the 
particle angular momentum and the canonical 
angular momentum are the same,
which leads to an incorrect and non-gauge invariant angular momentum
equation. (2) The electric and magnetic fields are not included
dynamically, as they must be for a plasma. 
We show that when the  Hamiltonian analysis is performed correctly, no 
new conclusions can be made out of it that cannot be made 
by writing down usual Lorentz forces and Maxwell equations.
Unfortunately, this exclude Hamiltonian analysis as a tool to 
resolve the problem of the existence of ADAF.
We hope that a broader consequence of our consideration will
help to elucidate the role and limitations of the Hamiltonian 
formalism in the context of such plasma dynamics problems.

\sectionb{2}{ANNOTATED DISCUSSION OF THE HAMILTONIAN\\ APPROACH 
OF BINNEY (2003)}
\label{sec2}

\subsectionb{2.1}{The Rate of Change of the Hamiltonian}

Binney (2003) uses a Hamiltonian formalism to calculate the change 
of energy and angular momentum of a particle under the action of 
electromagnetic and gravitational fields.  For a given time-dependent 
electromagnetic field characterised by the potentials $\vec{A}$ and $\psi$, 
and gravitational potential $\Phi$, the Hamiltonian for the non-relativistic 
motion of a particle of mass $m$ and charge $q$ in Cartesian coordinates 
is (e.g., Landau \& Lifshitz 1988b)
\begin{equation}
H=\frac{(\vec{p}-q{\vec{A}})^2}{2m}+ q\psi + m\Phi
\label{eq_1}\mbox{,}
\end{equation}
where $\Phi$ is assumed to be time-independent and axially 
symmetric, and the canonical momentum of the particle, $\vec{p}$, is 
related to its velocity $\vec{v}$ by
\begin{equation}
\vec{p}=m\vec{v}+q\vec{A}\mbox{,} 
\label{eq_2}
\end{equation}
where we  set the speed of light
$c=1$ to match the notation of Binney (2003).
Binney (2003) then calculates the time derivative of $H$. From 
general Hamiltonian theory it is known that the total time derivative
of any quantity $F$ along the trajectory of a particle is given by
\begin{equation}
{\rmd F\over \rmd t} = {\partial F\over \partial t} + \{ H,F \}, 
\label{eq_2a}
\end{equation}
where the Poisson bracket for position and momenta canonical variables 
$s_i,p_i$ is 
$\{H,F\}\equiv {\partial F\over \partial s_i}{\partial H\over \partial p_i}
-{\partial F\over \partial p_i}{\partial H\over \partial s_i}$
with repeated indices summed, 
and  ${\partial F/ \partial t}$ is the time derivative 
when $s_i,p_i$ are held fixed (Landau \& Lifshitz 1988a).
Since $\{H,H\}=0$, the rate 
of change of the Hamiltonian is 
\begin{eqnarray}
&& \frac{\rmd H}{\rmd t}=\frac{\partial H}{\partial t} = -q\frac{\vec{p} - q \vec{A}}
{m} \cdot \frac{\partial \vec{A}}{\partial t} + q \frac{\partial \psi}{\partial
t}   \nonumber \\
&& =-q \vec{v} \cdot \frac{\partial \vec{A}}{\partial t} + 
q\frac{\partial \psi}{\partial t}\label{dHdt} \mbox{.}
\label{4}
\end{eqnarray}

\subsectionb{2.2}{Hamiltonian is not the Particle Energy}

Binney (2003) then proceeds to identify $\rmd H/\rmd t$ from 
equation~(\ref{dHdt}) as a rate of change of the energy of a 
particle. He considers only the first term in the right hand side
of expression~(\ref{dHdt}) to obtain the lower limit on the rate 
of energy transfer, and the upper limit on the equipartition time
\begin{equation}
t_{\rm equi} \sim \frac{H}{|\rmd H/\rmd t|} < \frac{H}{| q \vec{v} \cdot 
\partial \vec{A}/\partial t |} \label{t_equi} \mbox{.}
\label{5}
\end{equation}
These expressions need to be summed over the particles in some 
small volume to obtain the corresponding rates for the plasma as 
a whole. Assuming that individual positive and negative charges
have the same charge magnitude $|q|$, we introduce the current density 
$\vec{j} =  |q| \sum_{l,m}(\vec{v}^{+}_{l} - \vec{v}^{-}_{m})$, where the sum is over
ions and electrons in some local volume of plasma where $\vec{v}_l^{+}$
is the velocity of the $l$th positive charge, 
and $\vec{v}_m^{-}$ is the velocity
of the $m$th negative charge. The result of separately 
summing up  $H$ and $\rmd H
/\rmd t$ over the charges then leads to:
\begin{equation}
t_{\rm equi} < \frac{H^{\rm tot}}{|\int \rmd^3 {\bf x} \vec{j}\cdot 
\partial \vec{A}/\partial t |} \label{t_equi1}\mbox{.}
\label{6}
\end{equation}

We must pause at this juncture.
Binney's (2003) presentation of (\ref{6}) as a measure of the time rate 
of change of the particle energy is flawed 
because $H$ is not in general a measure
of the particle energy. Specifically, note that the  
Hamiltonian of Eqn. (\ref{eq_1}) explicitly depends on time via $\vec{A}=
\vec{A}(\vec{r},t)$ and $\psi=\psi(\vec{r},t)$ 
and is not gauge invariant: the addition
of a time derivative of some scalar function to $\psi$ changes
$H$. Physical quantities such as the actual energy of a particle 
must be gauge invariant. 
It is not $H$ that represents the kinetic
energy of a particle, but rather it is  the first term in $H$ that
represents the particle kinetic energy. This term is 
explicitly gauge invariant (note that a gauge transformation also 
changes $\vec{p}$) due to its dependence on $\vec{A}$.
It is because of these points, that the identification 
of the  Hamiltonian with the energy of a particle and 
$\rmd H /\rmd t$ as the work done on a particle by the 
electric field cannot be correct. 
Rather, it is the kinetic energy of a particle, $K_{\rm e} = mv^2/2$,
that must be estimated to determine  equilibration 
times--the time derivative of the 
first term in Eq.~(\ref{eq_1}). 

The equipartition time
is not correctly obtained by (\ref{5}) but rather by  
$t_{\rm equi}= {1\over 2}m_{\rm e} v^2 / |\rmd K_{\rm e}/\rmd t| $, 
where $\rmd K_{\rm e}/\rmd t$ is simply
the time rate of change of the electron energy subject to acceleration
by interaction with the electromagnetic fields. It is clear without detailed
derivation that $\rmd K_{\rm e}/\rmd t$ is simply given by 
$q \vec{v}\cdot \vec{E}$, the work done per unit time by the
Lorentz force. Unlike in (\ref{5}) the correct expression for 
$\rmd K_{\rm e}/\rmd t$ will have $\vec{\nabla}\psi$ terms, which are
particularly dominant when close interactions of electrons and
ions occur (Coulomb collisions). Therefore, the correct estimate 
of $t_{\rm equi}$ is rather different from (\ref{5}) but is nothing new
compared to the estimate of the conventional work done by the electric field. 

Binney (2003) further argues  that in obtaining 
(\ref{t_equi}) the $q\partial\psi/\partial t$  terms of (\ref{4}) 
can be ignored because 
they cancel   when averaging over
particles with opposite charges in a quasi-neutral
plasma.  Actually, this cancelation is not guaranteed
because a particle with a given charge in a plasma always attracts 
particles of the opposite charge within its Debeye sphere. As a result, 
the change in the potential at the location of a particle 
correlates with its charge and the cancellation of 
$q\partial\psi/\partial t$ between species is not obvious.

We have identified problems with (\ref{t_equi1}) as presented
by Binney (2003), but let us for the moment, continue to follow and dissect 
the arguments that he presents in comparing the equilibration and infall
(angular momentum loss) 
times. We now discuss the latter and the pitfalls therein.

\subsectionb{2.3}{Canonical Angular Momentum is not the Particle\\ Angular Momentum\\}

To compare with the energy equilibration time (\ref{t_equi1}),
Binney (2003) proceeds to derive a  similar time scale for 
the loss of angular momentum of the plasma.  If the 
angular momentum loss time were the longer of the two,
he would rule out ADAFs because the electrons and protons would
equilibrate before the plasma falls onto the black hole. 
The fundamental problem with his subsequent calculation
is that it relies on incorrectly identifying the canonical
and particle angular momenta, as we now show.

Specifically, 
Binney (2003) identifies the angular momentum of a particle $L_z$ with
the $\phi$-component of the canonical momentum $p_{\phi}$. He uses 
Hamiltonian theory (e.g. Landau \& Lifshitz 1988a) to write
\begin{equation}
\frac{\rmd L_z}{\rmd t}=\{ H,p_{\phi}\} = -\frac{\partial H}{\partial \phi} = 
q\vec{v} \cdot \frac{\partial \vec{A}}{\partial \phi} - q\frac{\partial \psi}{\partial \phi}
\label{binney_dLzdt} 
\mbox{.}
\end{equation}
He then  notices that the right hand side of 
expression~(\ref{binney_dLzdt}) differs from the right hand side of 
expression~(\ref{dHdt}) by replacing $\partial t$ by $\partial \phi$ and 
changing sign. Therefore,  the analogous arguments and calculations
leading to the estimate~(\ref{t_equi1}) are repeated to obtain an estimate 
of the ``residence time'' or the maximum 
time before the plasmas loses its angular momentum and falls
into the black hole:
\begin{equation}
t_{\rm res} \sim \frac{L^{\rm tot}_z}{| \rmd L^{\rm tot}_z/\rmd t|} \sim
\frac{L^{\rm tot}_z}{| \int \rmd^3 \vec{x} \vec{j} \cdot \partial \vec{A} /\partial
\phi | } \label{binney_tres} \mbox{.}
\end{equation}
Dividing both sides of~(\ref{t_equi1}) by both sides of~(\ref{binney_tres}),
approximating $H^{\rm tot}/L^{\rm tot}_z$ by $\Omega_{\phi}$ (acceptable 
when the plasma is substantially supported by Keplerian rotation as 
in the case of ADAFs), and approximating the ratio of integrals by the ratio
of the corresponding derivatives of $\vec{A}$ one obtains
\begin{equation} 
\frac{t_{\rm equi}}{t_{\rm res}} < \Omega_{\phi} \frac{|\partial \vec{A}/\partial \phi|}
{|\partial \vec{A}/\partial t|}  \label{binney_ratio}\mbox{.}
\end{equation}
The ratio of derivatives of $\vec{A}$ can be approximated by the inverse 
frequency of a pattern of $\vec{A}$ propagating in $\phi$ direction.  
The lowest important frequency of the pattern in $\vec{A}$
is the Keplerian frequency $\Omega_{\phi}$.  Then, the final estimate follows:
\begin{equation}
\frac{t_{\rm equi}}{t_{\rm res}} < 1 \label{binney_final} \mbox{.}
\label{10}
\end{equation}
Taken at face value, this relation would imply that 
electrons receive a significant fraction of the thermal energy of ions
before they have time to fall through the black hole event horizon, and ADAFs 
are impossible.  We now pinpoint the key problem with the approach
that led to this conclusion.

\sectionb{3}{HAMILTONIAN EQUATIONS OF PLASMAS FROM\\ FIRST PRINCIPLES\\}

The key problem 
with the result and calculation that leads to (\ref{10})
above is the mis-identification of the angular momentum of a particle $L_z$
with the $\phi$-component of the canonical momentum 
$p_{\phi}$ above Eq. (\ref{binney_dLzdt}). To show that these are not the same and the important
consequences, we now rigourously derive $\rmd L_z/\rmd t$ starting from
first principles.

The forms of the Hamiltonian and canonical momenta depend on the choice 
of the coordinate system used to describe the particle motion. 
In cylindrical coordinates, expressions~(\ref{eq_1})
and~(\ref{eq_2}) need to be modified. The general procedure to derive 
Hamiltonian equations is to start with a Lagrangian (which can
be obtained from the covariant action). The Lagrangian
is invariant under the transformations of space coordinates $s_i$.
In cylindrical coordinates $(s_1, s_2, s_3)=$($r$,$\phi$,$z$) and the Lagrangian $\lag$
in the non-relativistic limit is
(e.g., Landau \& Lifshitz 1988b):
\begin{eqnarray}
&& \lag=\frac{m}{2}\left({\dot r}^2+r^2 {\dot\phi}^2 + {\dot z}^2\right)
+q\left(A_r {\dot r}+rA_{\phi}{\dot\phi}+A_z{\dot z}\right)
\nonumber\\
&& -q\psi-m\Phi\mbox{,}
\label{eq_3}
\end{eqnarray}
where ${\dot r}=\rmd r/\rmd t$, ${\dot \phi}=\rmd \phi/
\rmd t$, and ${\dot z}=\rmd z/\rmd t$ along the trajectory 
of a particle. The conjugate canonical momenta are $p_i=\partial \lag/\partial 
{\dot s_i}$. In cylindrical coordinates
\begin{equation}
p_r=m{\dot r}+qA_r, \> p_{\phi}=mr^2 {\dot\phi}+qrA_{\phi},
\> p_z=m{\dot z}+qA_z\mbox{.}
\label{eq_4}
\end{equation}
The actual particle angular momentum relative to the $z$-axis is 
$L_z=mr^2{\dot \phi}$. Eq.~(\ref{eq_4}) allows us to 
relate $L_z$ to $\phi$-component of conjugate momentum as 
\begin{equation}
L_z=p_{\phi}-qrA_{\phi}\mbox{.}\label{eq_5}
\end{equation}
Therefore, in electromagnetism $L_z \neq p_{\phi}$ contrary to the statement 
of Binney (2003). As a 
result, his equation~(\ref{binney_dLzdt})  for $\rmd L_z/\rmd t$ is incorrect. 

To derive the correct expression for $\rmd L_z/\rmd t$, 
we use the fact that the Hamiltonian $H=\sum_i {\dot s}_i \partial \lag/
\partial {\dot s}_i -\lag$. Then using~(\ref{eq_3}) and substituting for
${\dot s}_i$ from Eqs.~(\ref{eq_4}) gives
\begin{eqnarray}
&& H=\frac{1}{2m}\left[(p_r-qA_r)^2+(p_{\phi}/r-qA_{\phi})^2+
\right. \nonumber\\
&& \left. (p_z-qA_z)^2\right]+q\psi+m\Phi\mbox{.}\label{eq_6}
\end{eqnarray}
This expression is different from the expression~(\ref{eq_1})
for $H$ in Cartesian coordinates by the factor $1/r$ 
multiplied with $p_{\phi}$. Let us take 
$\rmd/\rmd t$ along the particle trajectory of all terms in 
relation~(\ref{eq_5}). One has
\begin{eqnarray}
\frac{\rmd p_{\phi}}{\rmd t}=\left\{H,p_{\phi}\right\}=
-\frac{\partial H}{\partial\phi} \nonumber \\
=q\left({\dot r}\frac{\partial A_r}{\partial\phi}+
r{\dot \phi}\frac{\partial A_{\phi}}{\partial\phi}+{\dot z}
\frac{\partial A_z}{\partial\phi}\right)
-q\frac{\partial\psi}{\partial\phi}\mbox{,}
\label{eq_7}
\end{eqnarray}
where the curly bracket is again the Poisson bracket and we have used
Eq.~(\ref{eq_4}).
This result 
exactly reproduces the right hand side of equation~(\ref{binney_dLzdt})  
but the left hand side should be 
$\rmd p_{\phi}/\rmd t$, {\bf not}
$\rmd L_z/\rmd t$.  There are extra terms to 
$\rmd L_z/\rmd t$ according to relation~(\ref{eq_5}):
\begin{eqnarray}
&& \frac{\rmd L_z}{\rmd t}=\frac{\rmd p_{\phi}}{\rmd t}
-q{\dot r}A_{\phi}-qr\frac{\partial A_{\phi}}{\partial t}-
qr\frac{\partial A_{\phi}}{\partial r}{\dot r}-
\nonumber\\
&& qr\frac{\partial A_{\phi}}{\partial \phi}{\dot \phi}-
qr\frac{\partial A_{\phi}}{\partial z}{\dot z}
\mbox{.}\label{eq_8}
\end{eqnarray}
Substituting $\displaystyle \frac{\rmd p_{\phi}}{\rmd t}$ from
Eq.~(\ref{eq_7}) into Eq. (\ref{eq_8})  
and collecting together terms with ${\dot r}$
and ${\dot z}$ we obtain
\begin{eqnarray}
&& \frac{\rmd L_z}{\rmd t}=q\left[r{\dot r}\left(\frac{1}{r}
\frac{\partial A_r}{\partial\phi}-\frac{1}{r}\frac{\partial}
{\partial r}(rA_{\phi})\right)+\right. \nonumber\\
&& \left. r{\dot z}\left(\frac{1}{r}
\frac{\partial A_z}{\partial\phi}-\frac{\partial A_{\phi}}
{\partial z}\right)\right] 
-rq\left(\frac{1}{r}\frac{\partial\psi}{\partial\phi}+
\frac{\partial A_{\phi}}{\partial t}\right)
\mbox{.}\label{eq_9}
\end{eqnarray}
When all components of potentials are combined in Eq.~(\ref{eq_9}),
we are left with  only components of magnetic and electric fields:
\begin{eqnarray}
&& \left(\frac{1}{r}
\frac{\partial A_r}{\partial\phi}-\frac{1}{r}\frac{\partial}
{\partial r}(rA_{\phi})\right)=-B_z, \nonumber \\
&& \left(\frac{1}{r}
\frac{\partial A_z}{\partial\phi}-\frac{\partial A_{\phi}}
{\partial z}\right)=B_r, \nonumber \\
&& -\frac{1}{r}\frac{\partial\psi}{\partial\phi}-
\frac{\partial A_{\phi}}{\partial t}=E_{\phi}.
\nonumber
\end{eqnarray}
Therefore, Eq.~(\ref{eq_9}) becomes
\begin{equation}
\frac{\rmd L_z}{\rmd t}=qrE_{\phi}+qr(v_z B_r -v_r B_z)
\mbox{.}\label{eq_10}
\end{equation}
This is nothing but the torque produced by the $\phi$ component
of the Lorentz force $q\vec{E}+q(\vec{v}\times\vec{B})$ 
acting on a particle. Indeed such a result could be anticipated
from the very beginning, without using the Hamiltonian
or Lagrangian formalism for the derivation.
In fact, equation~(\ref{binney_dLzdt}) after 
Binney (2003) would already raise initial concerns from the fact
that its left hand side 
$\rmd L_z/\rmd t$, is a gauge invariant quantity, while 
the right hand side is not gauge invariant.

The conclusion of Binney (2003) that we reproduced 
in the previous section culminating in 
Eq. (\ref{10}), 
crucially relies on 
the incorrect presence of the actual
rather than canonical 
momentum on the left side of Eq. (\ref{binney_dLzdt}). 
We have shown that when this equation~(\ref{binney_dLzdt}) is corrected,
no new results come from the Hamiltonian formalism
that do not already come from simply writing down the 
Lorentz force acting on a particle. 
The latter leads to classical estimates of equilibration times
used in standard two-temperature accretion flow calculations (e.g. Narayan
et al. 1998), and by contrast, the 
relation  expressed in Eq.({\ref{10}) is simply invalid.

\sectionb{4}{INCLUSION OF THE ELECTROMAGNETIC FIELDS\\ 
AS DYNAMICAL VARIABLES AND IMPLICATIONS FOR CALCULATING LOSS TIMES\\}
\label{sec3}

The energy transfer from protons to electrons in the turbulent accretion
plasma is mediated by electromagnetic fields (waves) 
excited by plasma instabilities. 
Particles and waves can exchange energy,
(see for example Begelman \& Chiueh (1988) in this accretion flow context).
In order to quantitatively assess
the effectiveness of the energy transfer between particles and
waves, one needs to couple the dynamical evolution 
of the electromagnetic fields  to dynamical evolution of the 
particle motion. This means that in general,  the electromagnetic
fields $\vec{E}$ and $\vec{B}$ (or $\vec{A}$ and $\psi$) 
need to be treated as dynamical variables, not
merely as background fields.
This contrasts Binney (2003), who  assumes that the electromagnetic field
is fixed:  $\vec{A}$ and $\psi$ are considered as  
externally imposed background fields on the dynamical system of particles,
not as dynamical variables themselves.

One might be  tempted 
to extend the intended Hamiltonian approach of Binney (2003) to  
include dynamical electromagnetic fields.
In such an approach, the electromagnetic fields can be decomposed
into the sum of normal modes and each such
mode could be treated as a dynamical degree of freedom. 
One could then write a Hamiltonian containing both 
particles and electromagnetic fields and use it to calculate 
the time derivatives of the total energy and angular momentum of 
the system. But the result of such calculation
is predetermined: the total energy and angular momentum of all particles 
and electromagnetic fields (including radiation) must be conserved
in time. Hamiltonian equations of motion written down for such a Hamiltonian
will simply be Maxwell equations for the electromagnetic fields combined with 
the equations of particles motion under the 
action of  Lorentz forces.
Such a procedure would not lead to  new results but
simply  lead us back to the conventional methods of analysing 
the equations of plasma dynamics which start with the equations of motion
in the first place.

\sectionb{5}{CONCLUSIONS}

We have elucidated the role of the 
Hamiltonian formalism in the study of plasmas
and its relation to the equations of motion. The latter is the usual
starting point for the practical 
study of plasma dynamics in applications to laboratory
and astrophysical plasmas. We have illustrated how 
the Hamiltonian formalism does not provide any more information than
that which is contained in the equations of motion when it comes to the 
practical calculation of dynamical time scales of a  system.

We have illustrated the importance of understanding these basic 
derivations by means of an application to a 
very current topic in astrophysics, namely, 
two-temperature, low luminosity accretion flows
commonly used as an explanation for the otherwise mysterious quiescent
accretion engines at the centres of galaxies.
In particular, we have discussed  Binney's (2003)
attempt  to use general Hamiltonian methods to obtain a constraint
on the ratio of  ion-electron equipartition time to
the angular momentum loss time of particles in these flows. 
If the former time scale were 
shorter than the latter, these accretion flows would be ruled out.

 We have shown  that the mathematically correct Hamiltonian
formalism does not provide any new information for estimating
the ion-electron equipartition time beyond  conventional
non-Hamiltonian approaches.
This is revealed when one corrects the Hamiltonian approach of Binney~(2003) 
by not equating the 
canonical angular momentum to the actual particle angular momentum.
The revised calculation shows that the  expression 
for angular momentum change of a particle used by Binney (2003)
is incorrect (evidenced also by the fact that it is 
not gauge-invariant) and thus the subsequent conclusion that 
the equilibration time between electrons
and ions in accretion plasma is always shorter than the accretion 
time is unsupported.  Instead, if performed correctly, Hamiltonian 
expressions for the rate of change of angular momentum of a particle
in electromagnetic fields lead to the usual 
torque provided by the Lorentz force.

We have also pointed out that to incorporate particle-wave
interactions occurring in turbulent plasmas,
one must treat the electromagnetic fields  as dynamical variables 
in the Hamiltonian formalism. 
Including these excitations will simply lead to a conserved Hamiltonian,
once again providing no new information beyond conventional
plasma physics approaches.

Finally, we note that Quataert (2003) also argued that the conclusion
of Binney (2003) is incorrect. Quataert (2003)
argued that the time scale on which
the energy of a particle changes due to the work by the electric field 
is not the time scale on which the true heating or change in entropy occurs.
He mentions two examples where this difference is evident: 
First is the motion of a particle in a slowly varying magnetic 
field, with characteristic variation time much longer than $\Omega^{-1}$,
where $\Omega=eB/mc$ is the cyclotron frequency. After some time the 
magnetic field returns to its initial value everywhere. In this case, 
$t_{\rm equi}$ calculated by the method of Binney (2003) 
[expression~(\ref{t_equi1}) above]  would be the 
characteristic variation time of the magnetic field. At the same time, 
in the absence of collisions, the energies of particles remain the same 
because of the conservation of the adiabatic invariant. 
His second example 
is an undamped Alfv\'en wave. In this wave the energy is transferred 
periodically between fields and particles, but there is no net heating.
As an extension of the argument about different time scales for adiabatic
and dissipative energy changes, Quataert (2003) mentions that particles
are heated at discrete wave-particle resonances,  not explicitly 
accounted for in Binney (2003).

Although these two examples of non-dissipative energy changes of particles
in time variable magnetic fields are clear 
and correct, the statistical nature of the turbulence 
(presumably existing in any accretion flow due to the non-linear development
of the magneto-rotational instability (MRI, e.g. Balbus \& Hawley 1998)) 
does not allow one to conclude  
that {\it all} particle-turbulence 
energy exchange processes will occur as in Quataert's two examples. 
Therefore, by themselves, these arguments of Quataert (2003) do not 
disprove the derivation of Binney (2003). In particular, Binney argued that 
the rate of heating may be estimated from 
equation~(2) in his paper [repeated as equation~(\ref{dHdt}) above in
this paper]  as follows:
``Thus this equation describes the mechanism by
which equipartition is established between ions and electrons; the net
direction of the energy flow is mandated by the general principles of
statistical physics, and the rate of flow may be estimated from
equation~(2).'' This statement does not contradict the specific energy
transfer examples
of Quataert described in our previous paragraph above.
The reason is that it is not clear how 
statistically important the examples of Quataert are for a 
realistic accretion flow. We have found different and more fundamental 
reasons that rigourously disprove the results of Binney (2003).

\vskip5mm

ACKNOWLEDGEMENTS. Support from DOE grant DE-FG02-00ER54600,
NSF grant AST-0406799, and NASA grant ATP04-0000-0016,
are acknowledged. 
VP also acknowledges partial support from the 
Centre for Magnetic Self-Organisation in Laboratory and Astrophysical
Plasmas funded by the National Science Foundation.

\References

\refb Balbus S. A., Hawley J. F. 1998,
Rev. Mod. Physics, 72, 1

\refb Begelman M. C., 
Chiueh T. 1988,  ApJ,  332, 872 

\refb Binney J.  2003, On the impossibility of advection dominated accretion.
Preprint, astro-ph/0308171

\refb Bisnovatyi-Kogan G. S., Lovelace R. V. E. 1997,  ApJ,
486, L43 

\refb Bisnovatyi-Kogan G. S., Lovelace R. V. E. 2001,
New Astronomy Review, 45,  663 

\refb Blackman E. G. 1999,  MNRAS,  302,  723 

\refb Gruzinov A. V. 1998, ApJ, 501,  787 

\refb Ichimaru S. 1977,  ApJ,  214,  840

\refb Krolik J. H. 1999, Active Galactic Nuclei : from the Central Black Hole 
to the Galactic Environment. Princeton University Press, Princeton

\refb Landau L. D., Lifshitz E. M. 1988a, 
Mechanics. Nauka, Moscow

\refb Landau L. D., Lifshitz E. M. 1988b, 
The Classical Theory of Fields. Nauka, Moscow

\refb Rees M. J., Phinney E. S., Begelman M. C.,  Blandford R. D. 
1982, Nature, 295, 17 

\refb Narayan R. 2002, in
Proc. Int. Conf. Lighthouses of the Universe: 
The Most Luminous Celestial Objects and Their Use for Cosmology,
ed. M Gilfanov  et al., 
European Southern Observatory, Garching, p. 405

\refb Narayan R., Mahadevan R., Quataert E. 1998, in Theory of Black Hole Accretion Disks, 
ed.  Marek A. Abramowicz, Gunnlaugur Bjornsson, and James E. Pringle,
Cambridge University Press, Cambridge, p.148

\refb Narayan R., Yi I. 1995, ApJ,  452, 710

\refb Narayan R., Igumenshchev I. V.,
Abramowicz M. A. 2000,  ApJ, 539,  798

\refb Quataert E. 1998, ApJ,  500,  978 

\refb Quataert E. 2003, On the viability of two-temperature accretion flows.
Preprint, astro-ph/0308451

\refb Quataert E., Gruzinov A. 2000, ApJ,  520,  248

\refb Shapiro S. L., Teukolsky S. A. 1983, Black Holes, White Dwarfs,
and Neutron Stars: The Physics of Compact Objects. 
Wiley-Interscience, New York

\end{document}